# CADe TOOLS FOR EARLY DETECTION OF BREAST CANCER

U. Bottigli[1], P.G. Cerello[2], P. Delogu[3], M.E. Fantacci[3], F. Fauci[4], G. Forni[5], B. Golosio[1], A. Lauria[5], E. Lopez[2], R. Magro[4], G.L. Masala[1], P. Oliva[1], R. Palmiero[5], G. Raso[4], A. Retico[3], S. Stumbo[1], S. Tangaro[6]

[1]Università di Sassari and Sezione INFN di Cagliari, Italy
[2]Sezione INFN di Torino, Italy
[3]Università and Sezione INFN di Pisa, Italy
[4]Università di Palermo and Sezione INFN di Catania, Italy
[5]Università "Federico II" and Sezione INFN di Napoli, Italy
[6]Università di Bari and Sezione INFN di Cagliari, Italy

**ABSTRACT**

A breast neoplasia is often marked by the presence of microcalcifications and massive lesions in the mammogram: hence the need for tools able to recognize such lesions at an early stage.
Our collaboration, among italian physicists and radiologists, has built a large distributed database of digitized mammographic images and has developed a Computer Aided Detection (CADe) system for the automatic analysis of mammographic images and installed it in some Italian hospitals by a GRID connection.
Regarding microcalcifications, in our CADe digital mammogram is divided into wide windows which are processed by a convolution filter; after a self-organizing map analyzes each window and produces 8 principal components which are used as input of a neural network (FFNN) able to classify the windows matched to a threshold.
Regarding massive lesions we select all important maximum intensity position and define the ROI radius. From each ROI found we extract the parameters which are used as input in a FFNN to distinguish between pathological and non-pathological ROI.
We present here a test of our CADe system, used as a second reader and a comparison with another (commercial) CADe system.

**INTRODUCTION**

Early diagnosis of breast cancer in asymptomatic women strongly reduces breast cancer mortality [1]. Screening programs, which consist in a mammographic examination performed for 49-69 years old women, is nowadays the best way to obtain this important aim. It has been estimated that screening programs radiologists fail to detect up to approximately 25% breast cancers visible on retrospective reviews and that this percentage increases if minimal signs are considered [2,3]. Sensitivity (percentage of pathologic images correctly classified) and specificity (percentage of non pathologic images correctly classified) of this examination increase if the images are analysed independently by two radiologists [4]. So independent double reading is now strongly recommended as it allows to reduce the rate of false negative examinations by 5-15% [5,6]. Recent technological progress has allowed to develop a number of Computer Aided Detection (CADe) systems [7], which can provide an automated detection of pathological strucures and act as a second reader to assist radiologists in diagnosis.

## DESCRIPTION OF A GPCALMA STATION

A breast neoplasia is often marked by the presence of microcalcifications and massive lesions in the mammogram, so we have developed a CADe system able to provide detection of these markers. Traditional mammograms can be digitised by means of a CCD linear scanner with a 85 µm pitch and 4096 gray levels. The images (18x24 cm$^2$) are stored in 10.5 Mbytes data files.

On these images our CADe is able to provide an automated detection for both microcalcifications and massive lesions (see fig. 1).

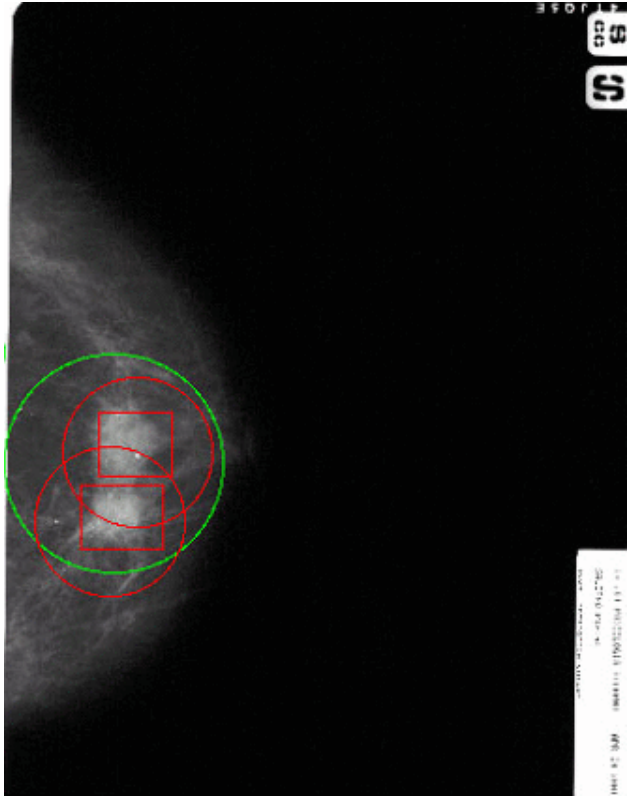

**Figure 1:** an example of the output of GPCALMA CADe. The green circle indicates the the ROI indicated by the radiologist as suspected for the presence of spiculated lesions with granular microcalcifications. The CADe has correctly indicated two ROIs suspect for the presence of pathological masses (red circles, threshold=0.9) and two ROIs suspect for the presence of microcalcifications (red rectangles, threshold=0.95). The histological examination has confirmed this detection, indicating the presence of a ductal infiltrating carcinoma with granular microcalcifications.

GPCALMA (Grid Platform for Computed Assisted Library for MAmmography) database is made of about 5500 distributed over various Italian hospitals. Different nodes will be connected using GRID technologies, allowing each node to work on the whole database.

## GPCALMA CADe SOFTWARE

### Opacities and Spiculated Lesions

Masses are rather large objects with very different shapes and faint contrast, slowly increasing with time. In GPCALMA database, the mean diameter of such lesions, as indicated by our radiologists, is 2.1cm. We have developed algorithms for recognition of opacities in general and specifically for spiculated lesions, which present a particular spike shape.

The interesting areas are selected by the construction of a structure with concentric rings centered on local intensity maxima until the mean pixel value reaches a fixed threshold, thus identifying ROIs consisting of circles of radius R. As a further step, for searching spiculated lesions, a spiral is unrolled around each maximum. For opacities, features are extracted by calculating the average intensity, variance and skewness (index of asymmetric distribution) of the pixel value distributions in circles of radius 1/3 R, 2/3 R and R, respectively. In the case of spiculated lesions, the number of oscillations per turn is calculated and processed by means of a Fourier Transform.

The features so extracted are used as input of a feed-forward neural network which perform the final classification. This network has an output neuron whose threshold (i.e. a number comprised between 0 and 1) represents the of suspiciousness of the corresponding ROI.

**Microcalcifications clusters**

A microcalcification is a rather small (0.1 to 1.0 mm in diameter) but very brilliant object. Some of them, either grouped in cluster or isolated, may indicate the presence of a cancer. In the GPCALMA database, the mean diameter of microcalcification clusters, as indicated by our radiologists, is 2.3cm.

Microcalcification cluster analysis is made using the following approach:
- digital mammogram is divided into 60x60 pixels wide overlapping windows;
- windows are statistically selected comparing the local and the global maxima;
- windows are shrunk from 60x60 to 7x7 and are classified (with or without microcalcifications clusters) using a FFNN with 49 input, 6 hidden, and 2 output neurons;
- windows are processed by a convolution filter to reduce the large structures;
- a self-organizing map (a Sanger's neural network) analyzes each window and produces 8 principal components;
- the principal components are used as input of a FFNN able to classify the windows matched to a threshold (the response of the output neuron of the neural network);
- windows are sorted by the threshold;
- at maximum three windows are memorized, if its threshold exceeds a given value;
- selected windows are zoomed to 180x180 pixels, i.e. 15x15 $mm^2$;
- overlapping windows are clusterized.

## COMPARISON WITH A COMMERCIAL CAD SYSTEM

Hence GPCALMA CADe is thoght to act as a second reader, we have tested its performances as an increase of radiologist detection performances, and we have compared them to radiologist detection improvement due to another CADe system: CADx Second Look, that is a commercial (FDA approved) CADe station.

A data set made of 190 (70 pathological, 120 without lesions) images has been shown to three different radiolologists (A, B, C), with a different degree of experience, A being the most expert and C a beginner. They made diagnosis on them in three different ways: without the support of any CADe system, supported by CADx and supported by GPCALMA CADe.

Results are presented in terms of sensitivity, that is the fraction of positive cases correctly detected to total positive cases, and specificity, that is the fraction of negative cases correctly detected to total negative cases

In tables 1 and 2 the results of the comparison of GPCALMA and Second Look, used as second readers, are shown.

**TABLE 1: Sensitivity (and confidence interval)**

|   | Alone (C.I.) | With CADx (C.I.) | With GPCALMA (C.I.) |
|---|---|---|---|
| **A** | 82.8% (4.5%) | 94.3% (2.8%) | 94.3% (2.8%) |
| **B** | 80.0% (4.8%) | 88.2% (3.8%) | 90.0% (3.6%) |
| **C** | 71.5% (5.4%) | 82.9% (4.5%) | 87.1% (4.0%) |

**TABLE 2: Specificity (and confidence interval)**

|   | Alone (C.I.) | With CADx (C.I.) | With GPCALMA (C.I.) |
|---|---|---|---|
| **A** | 87.5% (3.0%) | 84.2% (3.3%) | 87.5% (3.0%) |
| **B** | 91.7% (2.6%) | 85.9% (3.2%) | 88.4% (2.9%) |
| **C** | 74.2% (4.0%) | 70.8% (4.2%) | 70.9% (4.1%) |